\documentclass[a4paper,11pt]{article}
\usepackage{jheppub}

\pdfoutput=1
\usepackage[normalem]{ulem}
\usepackage{amsmath}
\usepackage{gensymb}
\usepackage{amsfonts}
\usepackage{comment}
\usepackage{amssymb}
\usepackage{mathrsfs}
\usepackage{graphicx}
\usepackage{color}
\usepackage{multirow}
\usepackage{bm}
\usepackage{tabularx}
\usepackage{subcaption}
\usepackage{blindtext}
\usepackage{wasysym}
\usepackage{makecell} 
\usepackage{float}
\usepackage{orcidlink}
\usepackage[normalem]{ulem}
\usepackage{lineno}
\usepackage[T1]{fontenc}
\usepackage{soul}
\usepackage{colortbl}
\usepackage{array}
\usepackage{cancel}
\usepackage[utf8]{inputenc}

\newcolumntype{L}{>{\centering\arraybackslash}m{1.99cm}}

\newcommand{\bi}{\begin{itemize}}
\newcommand{\ei}{\end{itemize}}

\begin{document}

\title{Solar Neutrinos at the ESSnuSB experiment
\\ \vspace{4mm}
{\small (ESSnuSB Collaboration)}}

\newcommand{\authorlist}{

\author[1]{J.~Aguilar,}
\author[2,*]{D.~Barčot, \note[*]{Corresponding authors}}
\author[3]{E.~Baussan,}
\author[4]{A.K.~Bhattacharyya,}
\author[5,6]{M.~Blennow,}
\author[7]{M.~Bogomilov,}
\author[4]{B.~Bolling,}
\author[3]{E.~Bouquerel,}
\author[8]{F.~Bramati},
\author[8,*]{A.~Branca,}
% ...

\affiliation[1]{Consorcio ESS-bilbao, Parque Científico y Tecnológico de Bizkaia, Laida Bidea, Edificio 207-B, 48160 Derio, Bizkaia, Spain}

\affiliation[2]{Center of Excellence for Advanced Materials and Sensing Devices, Ruđer Bo\v{s}kovi'c Institute, 10000 Zagreb, Croatia}

\affiliation[3]{IPHC, Universit'{e} de Strasbourg, CNRS/IN2P3, Strasbourg, France}

\affiliation[4]{European Spallation Source, Box 176, SE-221 00 Lund, Sweden}

\affiliation[5]{Department of Physics, School of Engineering Sciences, KTH Royal Institute of Technology,Roslagstullsbacken 21, 106 91 Stockholm, Sweden}

\affiliation[6]{The Oskar Klein Centre, AlbaNova University Center, Roslagstullsbacken 21, 106 91 Stockholm, Sweden}

\affiliation[7]{Sofia University St. Kliment Ohridski, Faculty of Physics, 1164 Sofia, Bulgaria}

\affiliation[8]{University of Milano-Bicocca and INFN Sez. di Milano-Bicocca, 20126 Milano, Italy}

% ...

} 
\emailAdd{dbarcot@irb.hr}
\emailAdd{mghosh@irb.hr}
\emailAdd{alessio.giarnetti@uniroma3.it}
\emailAdd{davide.meloni@uniroma3.it}

\authorlist

\abstract{We investigate the sensitivity of the ESSnuSB large underground water-Cherenkov (WC) detector to two key solar neutrino signatures: the MSW spectral upturn in the $^8$B electron neutrino survival probability $P_{ee}(E_\nu)$, and the first direct observation of the rare hep neutrino flux. The analysis incorporates detector systematic uncertainties (energy resolution and flux normalization) and models environmental backgrounds from cosmic-ray muon spallation and $^{222}$Rn contamination. We adopt a shape-driven analysis strategy with a floating overall normalization, so that the discovery power relies on the spectral distortion rather than on the absolute event rate. In a realistic scenario with 90\% spallation rejection efficiency, informed by the Super-Kamiokande experience and recent JUNO analyses, residual spallation backgrounds prevent either signal from reaching $5\sigma$ significance within 20 years of exposure. For the hep flux, we present a detailed sensitivity study to both its discovery and its precision, and quantify how the required spallation veto efficiency and recoil energy threshold jointly determine the achievable significance for both signals. At the adopted thresholds, discovery-level sensitivity for both signals requires approximately 99\% spallation rejection over a 20-year exposure.}

\maketitle

\section{Introduction}
\label{sec:intro}

The field of solar neutrino physics has a rich history spanning over several decades, playing a foundational role in our understanding of both stellar astrophysics and fundamental particle physics. The pioneering Homestake experiment~\cite{Davis:1968cp}, followed by real-time Water Cherenkov (WC)measurements at Kamiokande~\cite{Kamiokande-II:1989hgc}, revealed a substantial deficit in the observed flux of solar electron neutrinos compared to the robust predictions of the Standard Solar Model (SSM). This persistent discrepancy, famously known as the ``Solar Neutrino Problem'', was definitively resolved by the Sudbury Neutrino Observatory (SNO)~\cite{SNO:2001kky,SNO:2002tuh}. By measuring both the electron neutrino flux via charged-current interactions (CC) and the total active neutrino flux via neutral-current interactions (NC) of deuterium, SNO provided evidence of active neutrino flavor transformation during their travel from the solar core to the Earth.

This flavor-conversion mechanism is described by the Mikheyev-Smirnov-Wolfenstein (MSW) effect~\cite{Wolfenstein:1977ue,Mikheev:1986gs}, where neutrino propagation is modified by coherent forward scattering of electrons in dense matter. The Large Mixing Angle (LMA) MSW solution was subsequently confirmed with high precision by the KamLAND experiment~\cite{KamLAND:2002bdf}, which observed the disappearance of reactor antineutrinos over long baselines and firmly constrained the relevant mass-squared difference $\Delta m^2_{21}$ and mixing angle $\theta_{12}$. More recently, spectroscopic measurements by the Borexino experiment~\cite{Borexino:2014gjj, BOREXINO:2018ohr} have successfully mapped the individual components of the $pp$ chain fusion cycle, directly probing the transition between the low-energy vacuum-dominated regime and the high-energy matter-dominated regime. 

Despite these results, several details within the solar neutrino sector remain unverified, offering potential window into physics beyond the Standard Model (BSM). For instance, a long-standing $\sim 2\sigma$ tension exists between the value of $\Delta m^2_{21}$ extracted from solar neutrino data and that measured by KamLAND using reactor antineutrinos \cite{KamLAND:2013rgu,Xu:2022wcq}. Additionally, non-standard interactions (NSI) or light sterile neutrinos could cause anomalous distortions in solar electron neutrino survival probability $P_{ee}(E_\nu)$ \cite{Friedland:2004pp,Palazzo:2011vg,Chen:2022zts,Denton:2024upc,deHolanda:2003tx}. To rigorously test the standard MSW paradigm and search for such sub--leading BSM effects, it is essential to explore the transition region of the $^8$B neutrino spectrum. Within the standard LMA-MSW framework, $P_{ee}(E_\nu)$ is expected to transition from a low value of
\begin{equation}
  P_{ee}^{\rm HE} \simeq \sin^2\theta_{12}
  \label{eq:Phe}
\end{equation}
at high energies ($E_\nu \gtrsim 5$ MeV) to a higher value of
\begin{equation}
  P_{ee}^{\rm LO} = \tfrac{1}{2}\left(1+\cos^2 2\theta_{12}\right)
  \label{eq:Plo}
\end{equation}
at lower energies. This transition creates a characteristic spectral upturn in the intermediate energy range. While global fits strongly prefer this feature, it has not yet been directly observed with high significance in the $^8$B spectrum by the current experiments~\cite{Mollenberg:2014mfa,Hyper-Kamiokande:2018ofw,Barger:2005mn,Denton:2024upc,Smirnov:2006bu}. 

Simultaneously, the high-energy endpoint of the solar neutrino spectrum offers another astrophysics target: the $hep$ reaction ($^3$He $+$ $p \to$ $^4$He $+$ $e^+$ $+$ $\nu_e$). While the dominant $^8$B neutrino flux provides a window into the core temperature of the Sun, the rare $hep$ process produces neutrinos extending up to $\sim 19$ MeV, probing the outer layers of the solar core and the local distribution of $^3$He~\cite{Bahcall:2004pz,Vinyoles:2016djt}. According to SSM, $hep$ flux is more than two orders of magnitude smaller than $^8$B flux and has eluded direct observation, with only upper limits established by SNO and Super-Kamiokande (SK)~\cite{SNO:2020gqd,Super-Kamiokande:2023jbt}. A successful measurement of the $hep$ flux would complete our experimental mapping of the $pp$ chain and act as a highly sensitive test of solar interior dynamics. 

Another interesting effect observable with solar neutrinos, is the day-night asymmetry. It arises as solar neutrinos pass through the Earth, where interactions with dense terrestrial matter induce subtle flavor regeneration via the MSW effect. Consequently, measuring higher nighttime fluxes compared to daytime fluxes provides a direct experimental signature of terrestrial matter-enhanced oscillations \cite{Super-Kamiokande:2013mie,SNO:2005oxr,Xu:2022wcq}

The European Spallation Source Neutrino Super Beam (ESSnuSB)  is an upcoming long-baseline experiment that will use a high-intensity neutrino beam and a large underground WC detector to measure leptonic CP violation with high sensitivity~\cite{Alekou:2022emd,ESSnuSB:2023ogw}. To achieve this goal, the project will construct two massive 270 kt underground WC detectors. Due to its exceptional fiducial mass, deep underground location (1 km), and state-of-the-art photodetector layout, this facility offers an outstanding, high-statistics environment for non-beam neutrino physics \cite{ESSnuSB:2024wet,ESSnuSB:2025vsf,ESSnuSB:2026dar,ESSnuSB:2026ndb}, running concurrently with the main long-baseline accelerator program.

In this preliminary work, we investigate the sensitivity of the proposed ESSnuSB WC detector to two major low-energy solar neutrino signatures: the MSW-induced spectral upturn of $^8$B neutrinos and the first direct discovery of the $hep$ flux. Prior to dedicated Monte-Carlo simulations which will provide a precise evaluation of the detector response, we made realistic assumptions on the detector performances at low-energies. The paper is organized as follows: Section~\ref{sec:solar} reviews the solar neutrino interaction channels and background modeling at ESSnuSB; Section~\ref{sec:upturn} provides a quantitative assessment of the spectral upturn discovery potential; Section~\ref{sec:hep} analyses the sensitivity to the $hep$ neutrino flux while Section~\ref{sect:concl} summarizes our main conclusions and outlook. Finally in Appendix~\ref{app:dn} the day-night asymmetry at ESSnuSB is briefly discussed. 

\section{Solar Neutrinos at ESSnuSB}
\label{sec:solar}

The Sun produces neutrinos through several branches of the $pp$ chain and the CNO cycle, each with a characteristic energy spectrum: the dominant $pp$ neutrinos below 0.42 MeV, the mono-energetic $^7$Be line at 0.86 MeV, the $pep$ line at 1.44 MeV, and the higher-energy $^8$B and $hep$ continua extending up to $\sim15$ and $\sim18.8$ MeV, respectively. ESSnuSB will detect solar neutrinos through $\nu_e$--$e^-$ elastic scattering (ES) on the electrons of the water target. Because oscillated $\nu_\mu$ and $\nu_\tau$ interact only through the NC, the detected rate is dominated by the surviving electron-neutrino component, weighted by the relevant cross section. The recoil-electron spectrum produced by ES is a smeared, degraded version of the parent neutrino spectrum, and a detection threshold of a few MeV in recoil energy is unavoidable given the radioactive and cosmogenic backgrounds present in any large WC detector (see below). This threshold, together with the low endpoints of the $pp$, $^7$Be and $pep$ lines, means that ESSnuSB is only sensitive to the high-energy tail of the solar spectrum. In practice, only the $^8$B and $hep$ components produce recoil electrons above threshold; therefore, these are the two channels considered in this work. 

\subsection{Detector model and event rate}

We modeled the ESSnuSB detector as a cylindrical tank of 38~m radius and 76~m height filled with water. Applying a fiducial volume cut yields a fiducial mass of 540~kt for the combined two tank configuration, which we adopt throughout this work \cite{ESSnuSB:2023ogw}. 
The number of target electrons is computed in the following way:
\begin{equation}
    N_e = (\rho \pi R^2 H / M_{\rm H_2O})\cdot N_A \cdot 10
\end{equation} 
where $\rho$ is the density of water, $R$ and $H$ are respectively the radius and height of the cylindrical fiducial volume, $M_{\mathrm{H_2O}}$ is the molar mass of water and $N_A$ is the Avogardo number. The final factor of (10) accounts for the number of electrons in each neutral water molecule. For the ES cross section, we use the tree-level expression of Ref.~\cite{Bahcall:1995mm},
\begin{equation}
\frac{d\sigma}{dT}(E_\nu,T) = \frac{2G_F^2 m_e}{\pi}
\left[g_L^2 + g_R^2\!\left(1-\frac{T}{E_\nu}\right)^2 - g_L g_R \frac{m_e T}{E_\nu^2}\right],
\end{equation}
where $E_\nu$ is the incident neutrino energy, $T$ is the kinetic energy of the recoiling
electron, $G_F$ is the Fermi constant, and $m_e$ is the electron mass.
The quantities $g_L$ and $g_R$ are the left- and
right-handed weak couplings. The recoil energy is then integrated from the detector threshold
$T_{\rm th}=5~\mathrm{MeV}$ up to the kinematic endpoint $T_{\rm max}(E_\nu)=2E_\nu^2/(m_e+2E_\nu)$.

Radiative corrections are known to reduce the total neutrino-electron scattering cross section by approximately 2\%~\cite{Bahcall:1995mm}. They also introduce an energy-dependent distortion of the $^8$B recoil-electron spectrum, reaching about 4\% at the highest recoil energies. More recent calculations have further investigated the energy dependence of these corrections, including QED photon radiation and its implications for the reconstructed recoil-electron spectrum~\cite{Tomalak:2019ibg,Huang:2024rfb}.
Although the overall normalization effect is partially absorbed in our shape-driven upturn analysis, the energy-dependent contribution may affect the predicted spectral shape. A quantitative evaluation of its impact  is left for future analyses.
The expected event rate per reconstructed energy bin $k$ is then
\begin{equation}
\mu_k = N_e\, t\, \Delta \bar{ E}\int dE_\nu\;
0.9\;\phi(E_\nu)\,P_{ee}(E_\nu)\,\sigma(E_\nu)\, R(E_\nu, E_{\rm rec}),
\label{eq:rate}
\end{equation}
where $t$ is the exposure in years, $\Delta \bar{E}$ is the spacing of the true-energy grid, $\phi(E_\nu)$ is the solar neutrino flux and $P_{ee}(E_\nu)$ is the electron neutrino survival probability (see next subsection). The factor $0.9$ accounts for the overall signal detection efficiency.
The detector response $R(E_\nu,E_{\rm rec})$ was taken to be a Gaussian smearing of fractional width $\sigma_E = f_{\rm res}\cdot E_{\nu}$, truncated at the kinematic endpoint. We adopt $f_{\rm res} = 10\%$ for the upturn analysis ($E_{\rm rec}\in[6.5,15]$ MeV) and $f_{\rm res}=5\%$ for the $hep$ analysis ($E_{\rm rec}\in[14,25]$ MeV), taking into account the improved fractional resolution expected at higher energy.

\subsection{Survival probability}

\begin{figure}[h]
    \centering
    \includegraphics[width=0.95\linewidth]{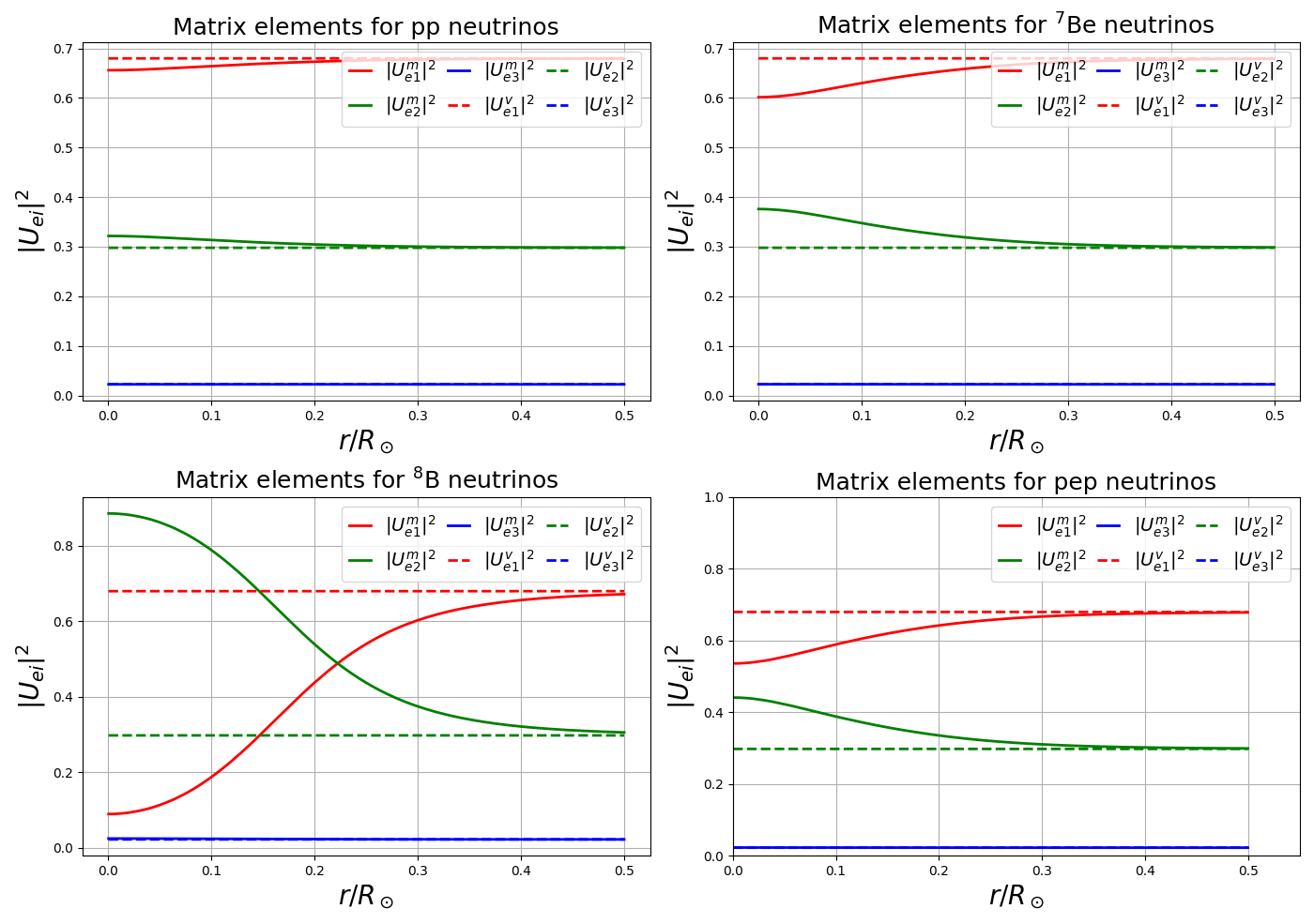}
    \caption{{Squared PMNS matrix elements in vacuum (dashed) and matter (solid) as a function of the neutrino production radius $r/R_\odot$ for the $pp$, $^7$Be, $^8$B, and $pep$ solar neutrino components. Only the $^8$B panel shows a sizeable matter/vacuum splitting across the production profile; the resulting radius-dependence of $P_{ee}$ is the origin of the spectral upturn.}}
    \label{fig:matrix_elements}
\end{figure}

The survival probability $P_{ee}(E_\nu)$ that enters Eq.~\eqref{eq:rate} is computed in the three-flavour adiabatic MSW approximation~\cite{Denton:2024upc},
\begin{equation}
\label{eq:pee}
P_{ee}^{(a)}(E_\nu) = \int dr\;\varphi_a(r)
\sum_{i=1}^3 |U_{ei}^{\nu}|^2 |U_{ei}^{m}(r,E_\nu)|^2,
\end{equation}

where $U_{ei}^{\nu}$ are the vacuum PMNS matrix elements, while $U_{ei}^{m}(r,E_\nu)$ are the corresponding matter-modified elements evaluated at the neutrino production point. The function $\varphi_a(r)$ is the normalized radial production distribution for the solar neutrino component $a$, satisfying $\int dr\,\varphi_a(r)=1$. Both the solar electron density profile used to compute $U_{ei}^{m}(r,E_\nu)$ and the production-radius distributions are obtained from the SSM~\cite{Bahcall:2004pz}. The energy dependence of $P_{ee}$ arises from the MSW matter potential, $A(r,E_\nu)=2\sqrt{2}G_F N_e(r)E_\nu$, where $N_e(r)$ is the solar electron number density. The energy-dependent solar neutrino flux enters separately in the event-rate calculation of Eq.~\eqref{eq:rate}. In the computation, $\theta_{12}=33.68^\circ$, $\Delta m^2_{21}=7.49\times10^{-5}$ eV$^2$ were used\footnote{Notice that the mass splitting $\Delta m_{21}^2$, which is not a parameter of the mixing matrix, enters in the oscillation probabilities when including the matter MSW effect \cite{Wolfenstein:1977ue}.}; the remaining oscillation parameters were taken from the NuFit-6.0 global fit~\cite{Esteban:2024eli}.
Fig.~\ref{fig:matrix_elements} illustrates how the squared PMNS matrix elements evolve from their vacuum values (dashed) to the matter-dominated values (solid) as a function of the production radius $r/R_\odot$, separately for the four solar components that populate the ESSnuSB energy window and below. The curves are obtained considering neutrino energy 0.826 MeV and 1.44 MeV for the $^{7}$Be and $pep$ lines, respectively; for the $pp$ and $^8$B continuous spectra the reference energies 0.267 MeV and 8.1 MeV have been used. For $pp$, $^7$Be and $pep$ neutrinos, the matter and vacuum mixing elements essentially coincide across the whole production region, since these neutrinos are produced and propagate at energies well below the MSW resonance; the corresponding survival probability is close to the vacuum-averaged value throughout. The $^8$B panel is qualitatively different: at small production radii, where the electron density is largest, $|U_{e1}^m|^2$ and $|U_{e2}^m|^2$ are far from their vacuum values and the propagation is fully matter-dominated, giving $P_{ee}\simeq\sin^2\theta_{12}$ at the highest energies; towards the outer, lower-density regions of the production profile the matter and vacuum curves converge. It is precisely this radius and energy dependent transition between the two regimes that produces the spectral upturn discussed in Sec.~\ref{sec:upturn}, appearing in the intermediate energy region, i.e. around 5 MeV, of the $^8$B spectrum where neither the pure matter, nor the pure vacuum limit is a good approximation.

\subsection{Backgrounds}
Two environmental backgrounds are relevant for the recoil-electron energy range considered here: $^{222}$Rn-chain radioactivity and cosmic-ray muon spallation products. 

\paragraph{Radon wall}
\label{par:radon}

The $\beta$-decay of $^{214}$Bi ($Q=3.27$ MeV) from the $^{222}$Rn chain produces a steeply falling background that we model as a Gaussian tail centered on $Q_{\rm Bi}$. The residual rate is normalized to the value measured by SK-IV, $9.53$ events/day/kt,  integrated over $3.5$--$5.0$~MeV, after SK's own radon-suppression cuts~\cite{Nakano_2017}. Because ESSnuSB has fewer photomultipliers per unit area than SK (30\%~\cite{ESSnuSB:2024wet} coverage vs.\ SK's 40\% \cite{Super-Kamiokande:2023jbt}),  it collects fewer photons per event, so it will reconstruct event positions less precisely, and a comparable radon-rejection cut will let more events through. We estimate this leakage (decays reconstructed on the wrong side of the fiducial cut) grows with the photon statistics as $1/\sqrt{N}$, giving a scale factor of  $\sqrt{0.40/0.30}\approx1.15$ relative to SK's residual rate. This contribution is entirely negligible above the 6.5 MeV analysis threshold adopted for the upturn study, and plays no role in the $hep$ analysis window. However, as shown in the left panel of Fig.~\ref{fig:upturn_char}, the smeared radon tail extends to nearly 5 MeV. It must therefore be included in the recoil-threshold scan discussed in Sec.~\ref{sec:upturn}, which examines the effects of lowering the recoil-electron energy threshold.

\paragraph{Muon spallation}
\label{par:spallation}
Cosmic-ray muons that interact with $^{16}$O instead produce a range of unstable spallation isotopes ($^{12}$B, $^{12}$N, $^{16}$N, $^8$Li, among others) that subsequently decay via $\beta/\beta\gamma$ emission~\cite{Super-Kamiokande:2015xra}, populating exactly the energy range of interest for both the $^8$B and $hep$ signals. Therefore, the spallation background is the sum of several distinct isotopes, each with its own production rate and $\beta$-decay endpoint, and in Secs.~\ref{sec:upturn} and \ref{sec:hep} we make use of a detailed, isotope-resolved model to study how the required veto efficiency and recoil threshold affect the sensitivity.  Isotope production rates are measured by SK in Kamioka~\cite{Super-Kamiokande:2015xra}. While both Kamioka and the planned ESSnuSB location in Zinkgruvan mine sit at $\sim$1~km depth\footnote{Usually quoted as 2700~mwe (meters water equivalent); the standard way of expressing overburden as the depth of water that would give the same shielding.}, Zinkgruvan has a flat overburden, while Kamioka has a mountain. This is important because a mountain lets in more muons than a flat rock at the same vertical depth. We rescale SK's isotope production rates, using the cosmic muon depth-intensity relation of Mei \& Hime~\cite{Mei_2006}: 
\begin{equation}
  I(h) = 67.97\times10^{-6}\, e^{-h/0.285} + 2.071\times10^{-6}\, e^{-h/0.698}
  \ \ [\mathrm{cm^{-2}s^{-1}}]
  \label{eq:burden_conversion}
\end{equation}
evaluated at Zinkgruvan's depth $h_{\rm ESS}\approx2700$~mwe, with $I_{\rm SK} = 1.49\times10^{-7}~\mathrm{cm^{-2}s^{-1}}$ directly measured at Kamioka~\cite{Abe_2010}.  We take the ratio
$R_{\rm ESS/SK} = I(h_{\rm ESS})/I_{\rm SK}$ between this predicted ESSnuSB flux and SK's measured flux, and use it to rescale SK's per-isotope spallation production rates to ESSnuSB, assuming isotope production scales linearly with muon flux.

In all cases, we consider three scenarios for the residual spallation rate, namely the background rate after the rejection cut: \textit{ideal} (no environmental background), \textit{realistic} (90\% rejection efficiency, motivated by the SK spallation-rejection performance~\cite{Super-Kamiokande:2023jbt} and the recent JUNO solar-neutrino spallation analysis~\cite{JUNO:2025spall}), and \textit{pessimistic} (no veto applied).

\section{Spectral Upturn Discovery}
\label{sec:upturn}

\begin{figure}[h]
    \centering
    \begin{minipage}[t]{0.49\linewidth}
        \centering
        \includegraphics[width=\linewidth, height=5.5cm]{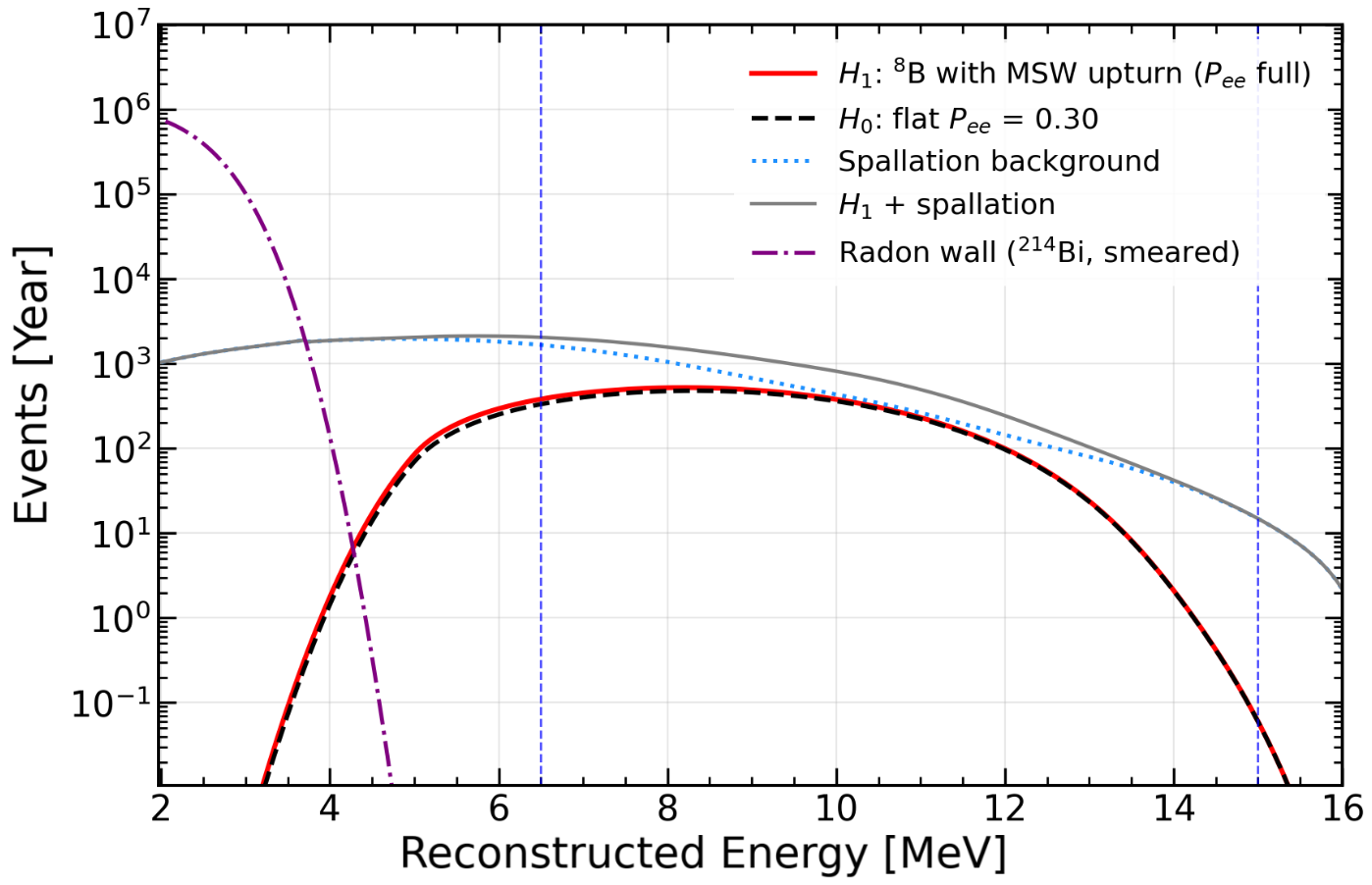}
    \end{minipage}\hfill
    \begin{minipage}[t]{0.46\linewidth}
        \centering
        \raisebox{0.5mm}{%
        \includegraphics[width=\linewidth, height=5.5cm]{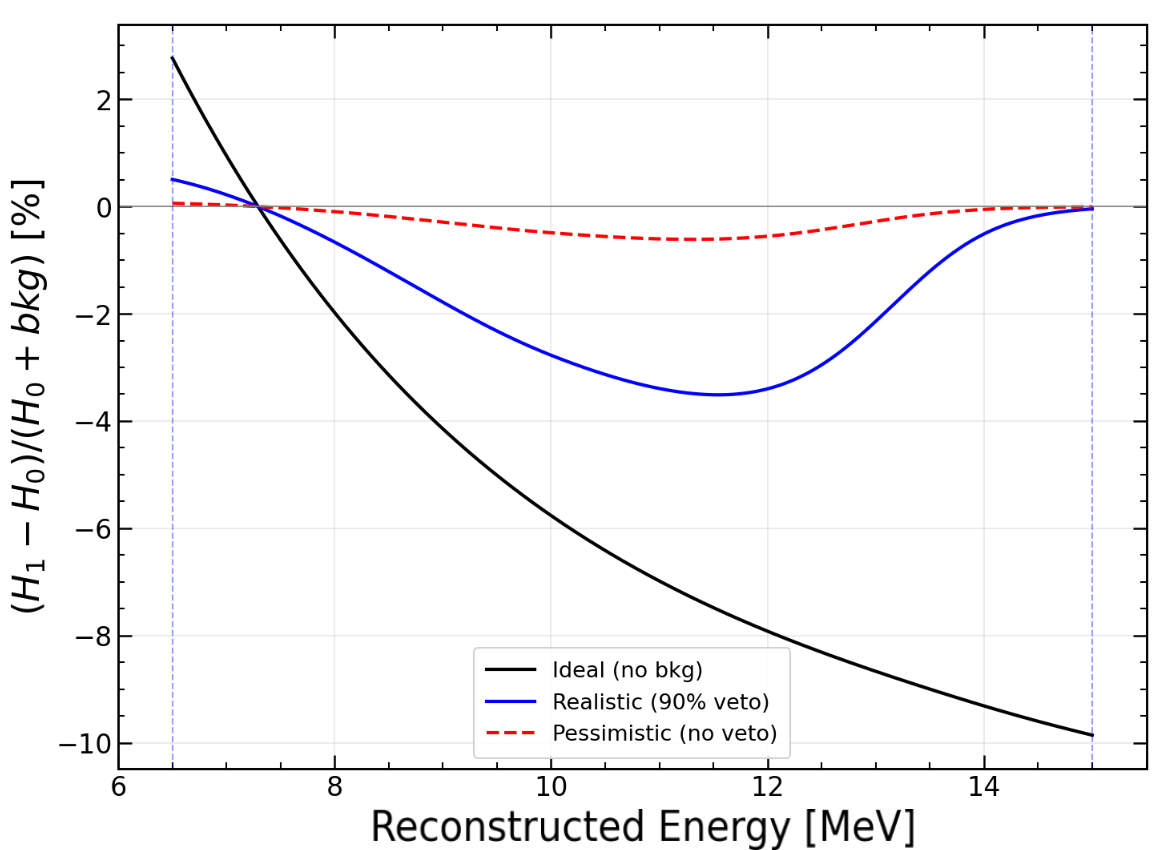}
        }
    \end{minipage}
    \caption{\textit{Left}: reconstructed energy spectra used in the upturn analysis: the $H_1$ hypothesis ($^8$B with the full MSW upturn, red solid), the $H_0$ hypothesis (flat $P_{ee}=0.30$, black dashed, used for the illustration purposes), the muon-spallation background (blue dotted), sum of the count of H1 hypothesis and spallation background (grey), and the $^{214}$Bi Radon-wall contamination (purple dash-dot); the vertical dashed lines delimit the $[6.5,15]$ MeV analysis window. \textit{Right}: fractional spectral difference $(H_1-H_0)/(H_0+{\rm bkg})$ between the upturn and flat hypotheses, for the ideal (no background, black), realistic (90\% veto, blue) and pessimistic (no veto, red dashed) scenarios. }
    \label{fig:upturn_char}
\end{figure}

We test the full MSW oscillation prediction ($H_1$: $P_{ee}(E_\nu)=P_{ee}^{\rm MSW\text{-}LMA} \equiv P_{MSW}$) against a flat survival probability. The analysis window is $E_{\rm rec}\in[6.5,15]$ MeV in 1 MeV bins; the lower cut removes the contamination of $^{214}$Bi discussed above. Since the MSW transition is centered around $E_\nu\sim5$~MeV (Sec.~\ref{sec:intro}), the analysis is primarily sensitive to the higher energy region of the upturn. Fig.~\ref{fig:upturn_char} summarizes the spectral information entering the analysis. The left panel shows the reconstructed spectra used in the comparison: the $H_1$ hypothesis ($^8$B with the full MSW upturn), the $H_0$ hypothesis (flat $P_{ee}$), the muon-spallation background components with their sum, and the residual $^{214}$Bi Radon-wall contamination; the two hypotheses are essentially indistinguishable by eye in the presence of the spallation background, which is more than an order of magnitude above the signal near threshold. This is the origin of the strong dependence of the upturn sensitivity on the spallation veto efficiency discussed below, and it is quantified directly in the right panel, which shows the fractional difference $(H_1-H_0)/(H_0+{\rm bkg})$ as a function of reconstructed energy for the ideal (no background), realistic (90\% veto) and pessimistic (no veto) scenarios. In the background-free case, the distortion grows monotonically towards high energy, reaching nearly $-10\%$ at the high-energy edge of the window and changing sign around 7.3 MeV; This behaviour follows directly from the energy dependence of $P_{ee}$: since $P_{\mathrm{MSW}}$ decreases monotonically with energy, from the vacuum limit at low energies to the matter limit near 15~MeV, $H_1$ predicts more events than the best-fit flat $H_0$ at low energies, where $P_{\mathrm{MSW}} > P_{\mathrm{flat}}^{\mathrm{best\text{-}fit}}$, and fewer at high energies, where $P_{\mathrm{MSW}} < P_{\mathrm{flat}}^{\mathrm{best\text{-}fit}}$. Note that this best-fit flat $P_{ee}$ is not the illustrative value of $0.30$ shown in the left panel, but is instead left free to float within its physical bounds $[0.294,\,0.553]$ (the vacuum and matter asymptotes); it is this floating best fit, rather than a fixed reference value, that produces the sign change seen on the right. Although the absolute size of the signal does not grow towards high energy, the plotted quantity does, because it is normalized to the prediction $H_0$: the denominator falls steeply with energy, inflating an otherwise small residual into a large fractional deviation, and this growth drives the sensitivity at high energy. Including the spallation background leaves the numerator unchanged, since the background is identical under $H_1$ and $H_0$ and cancels in the difference, but it adds to the denominator - diluting the fractional difference more strongly where the background is larger relative to the signal, which is why with the realistic (90\% veto) and pessimistic (no veto) the shape difference is diluted and develops a markedly different energy dependence. The realistic (90\% veto) curve showing a minimum of about $-3.5\%$ near 11--12 MeV before rising again, while the pessimistic (no-veto) curve stays within about half a percent of zero across the whole window. This illustrates directly why the residual spallation background is the dominant limiting factor for the upturn measurement: it not only adds statistical noise but also reshapes the very feature that the analysis is designed to detect.

For a set of reconstructed-energy bins, we construct an Asimov dataset, in which the event counts $n_k$ are set equal to their expected values under the $H_1$ (MSW-LMA) hypothesis, without statistical fluctuations. The quantity $\mu_k(\theta)$ denotes the predicted event counts in bin $k$ under the $H_0$ (flat survival probability) hypothesis, including signal and background contributions, while $\theta$ represents the parameters varied in the fit, namely the constant survival probability $P_{\rm flat}$ and the nuisance parameters associated with energy resolution and background normalization. The test statistic is then defined as
\begin{equation}
  \chi^2(\theta) = (n-\mu)^T V^{-1}(\mu)\,(n-\mu), \qquad
  V = \mathrm{diag}(\mu) + \alpha^2\,\mu\mu^T,
  \label{eq:chi2}
\end{equation}
The first term of $V$ is the usual Poissonian statistical covariance, the second one is a fully-correlated rank-1 term with $\alpha = 10\%$, representing a coherent normalization uncertainty (as $^8$B flux or fiducial mass) that shifts every bin in the same direction. 
For a rate-only deviation, the correlated normalization uncertainty limits the achievable significance at large exposures, with the asymptotic significance scaling as the fractional rate difference divided by $\alpha$. A larger $\alpha$ therefore reduces the maximum achievable significance. Since $P_{\rm flat}$ is allowed to vary within its physical bounds, the fit can absorb part of the overall rate difference between $H_1$ and $H_0$, while the correlated covariance term further reduces sensitivity to coherent normalization shifts. The remaining sensitivity is therefore primarily driven by differences in the spectral shape. As a result, the upturn discovery significance is largely insensitive to the precise value of $\alpha$ within this shape-driven framework; the dominant residual systematic is the detector energy resolution, which can partially mimic the spectral distortion itself. A purely statistical covariance would overestimate the sensitivity to a coherent rate excess, which could always be absorbed
by a flux rescaling. The floating of the $P_{\rm flat}$ value is restricted to the physically motivated MSW range $P_{\rm flat} \in [0.294, 0.553]$. The bounds are the two asymptotic limits of $P_{ee}(E_\nu)$ (Eqs.~\eqref{eq:Phe}--\eqref{eq:Plo}), evaluated with the same oscillation parameters mentioned in Sec.\ref{sec:solar} and including the small $\cos^4\theta_{13}$ suppression from the third mixing angle: $0.294 = \sin^2\theta_{12}\cos^4\theta_{13}$ and $0.553 = \tfrac12(1 +\cos^22\theta_{12})\cos^4\theta_{13}$. Since the real energy-dependent $P_{ee}(E_\nu)$ is by construction confined between these two values at every energy, any single constant $P_{\rm flat}$ meant to represent "no upturn" must also be within this range. If a fit is allowed to push $P_{\rm flat}$ outside of the given range, that would no longer correspond to a physically meaningful flat probability. The correlated normalization uncertainty is incorporated through the covariance matrix $V$, while Gaussian penalty terms for the energy-resolution and background-normalization nuisance parameters are added to $\chi^2$ where relevant. The profiled significance is then $Z = \sqrt{q}$, with $q = \min_{\hat\theta}\chi^2(\theta)$, which by Wilks' theorem is
asymptotically a standard normal deviate.

\begin{figure}[h]
    \centering
    \begin{minipage}[t]{0.49\linewidth}
        \centering
        \includegraphics[width=\linewidth, height=5.5cm]{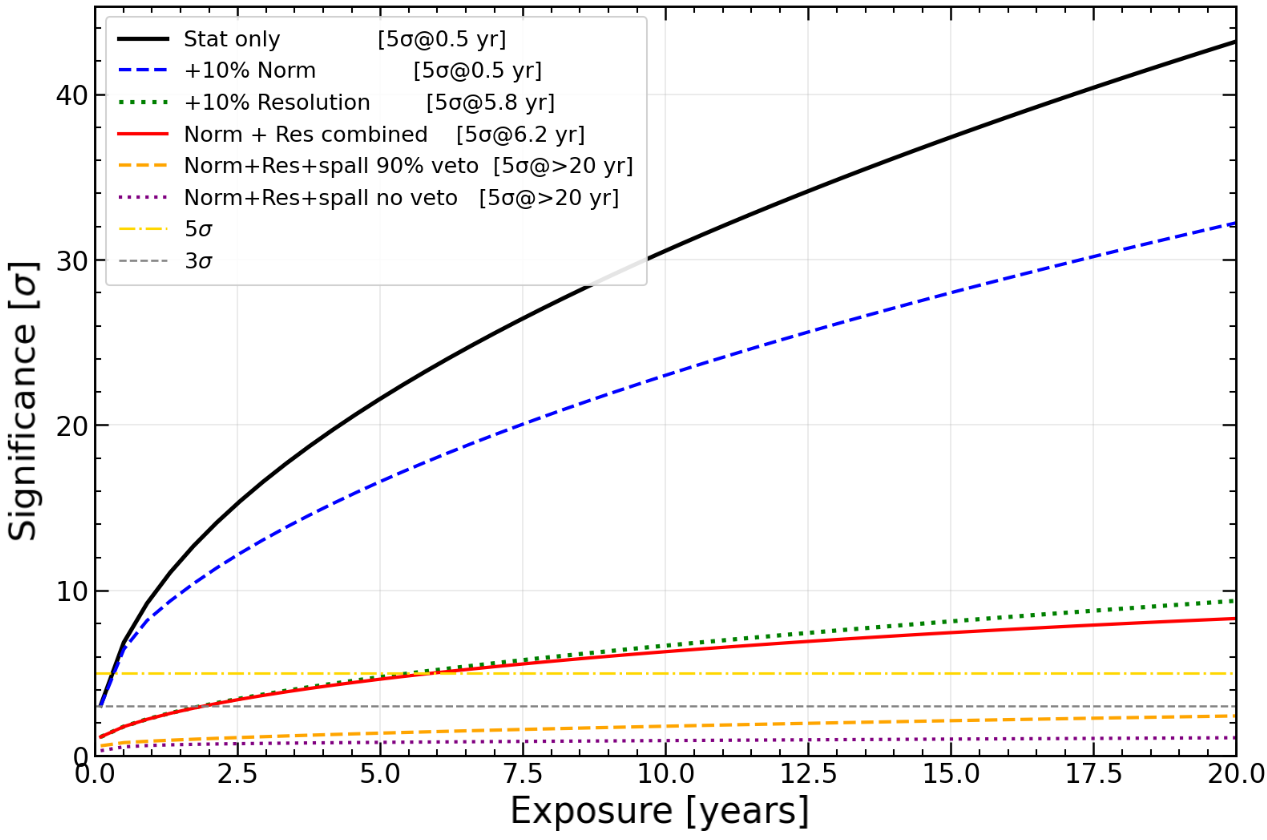}
    \end{minipage}\hfill
    \begin{minipage}[t]{0.49\linewidth}
        \centering
        \includegraphics[width=\linewidth,height=5.5cm]{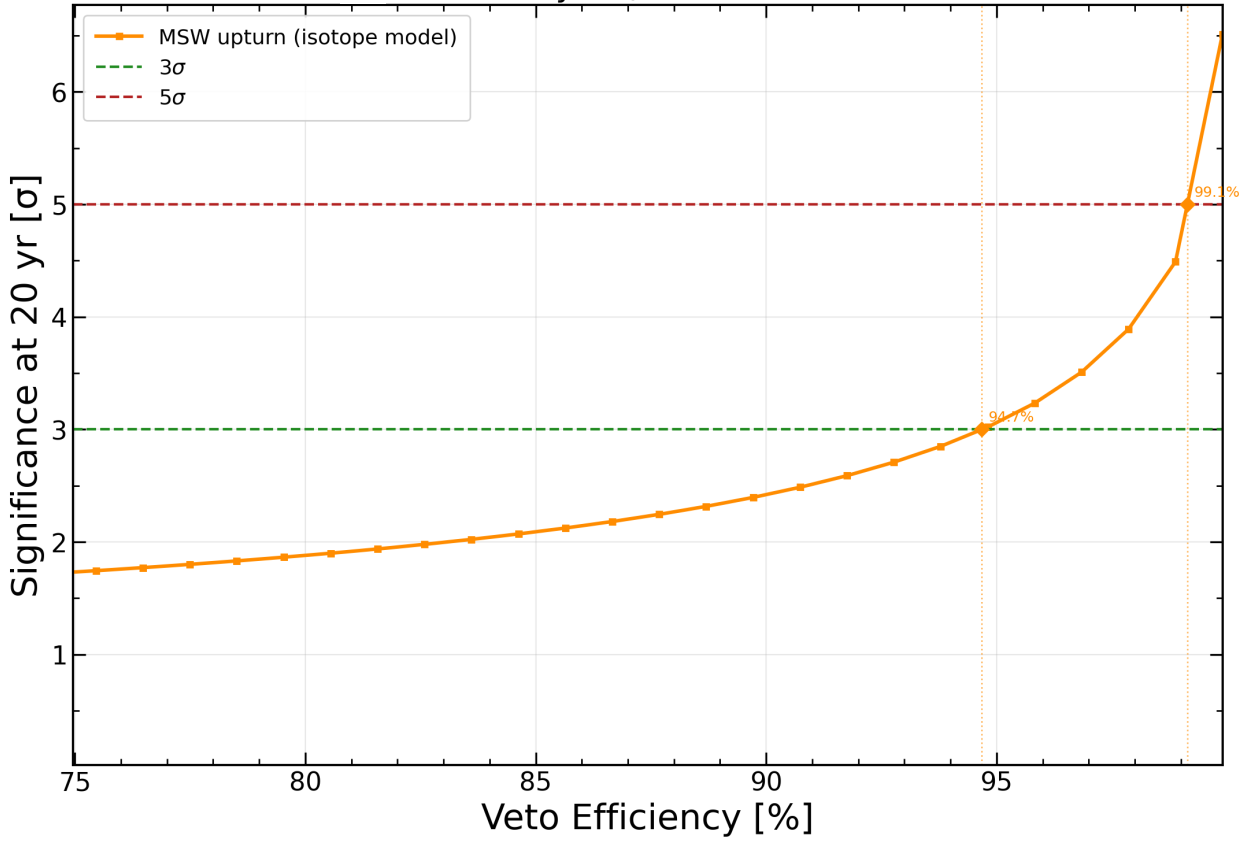}
    \end{minipage}
    \caption{\textit{Left}: upturn discovery significance vs.\ exposure: statistical only (black), $+10\%$ normalization (blue dashed), $+10\%$ resolution (green dotted), Norm+Res combined (red), and Norm+Res with a 90\% or no spallation veto (orange dashed, purple dotted). \textit{Right}: upturn discovery significance after 20 years of exposure, as a function of the spallation veto efficiency, using the isotope-resolved spallation model. The $3\sigma$ and $5\sigma$ reference lines are shown.}
    \label{fig:upturn_sens}
\end{figure}

Fig.~\ref{fig:upturn_sens} shows the resulting discovery significance. The left panel shows the significance as a function of exposure: in the purely statistical case (namely with no systematics and no backgrounds) $5\sigma$ is reached in about half a year; adding a 10\% normalization uncertainty on the $^{8}$B flux lowers the significance, which, however, still remain high enough to reach more than $5\sigma$ within one year. A 10\% resolution uncertainty alone pushes the $5\sigma$ threshold to $\sim 5.8$ years, while the combination of normalization and resolution uncertainties (Norm+Res) pushes the $5\sigma$ threshold to $\sim6.2$ years. Once the spallation background is included, even with a 90\% veto efficiency, the significance saturates well below $5\sigma$ over a 20-year exposure, and with no veto the sensitivity is essentially destroyed. The right panel shows, using the isotope-resolved spallation model introduced in Sec.~\ref{sec:solar}, the significance after 20 years as a continuous function of the spallation veto efficiency: the curve remains below $3\sigma$ up to a veto efficiency of $\sim94$--95\%, rising sharply thereafter and only crossing $5\sigma$ near $\sim99\%$, indicating that a discovery-level upturn measurement requires spallation rejection efficiencies well beyond the 90\% achieved by Super-Kamiokande, at least with the current 5~MeV threshold on electron recoil energy. The exposure of 20 years is chosen because the upturn signal is intrinsically small, so a correspondingly long exposure is needed to accumulate sufficient statistics for a discovery-level measurement; 20 years is adopted as a representative value and is used consistently throughout the paper.

\begin{figure}[h]
    \centering
    \begin{minipage}[t]{0.48\linewidth}
        \centering
        \includegraphics[width=\linewidth, height=5.5cm]{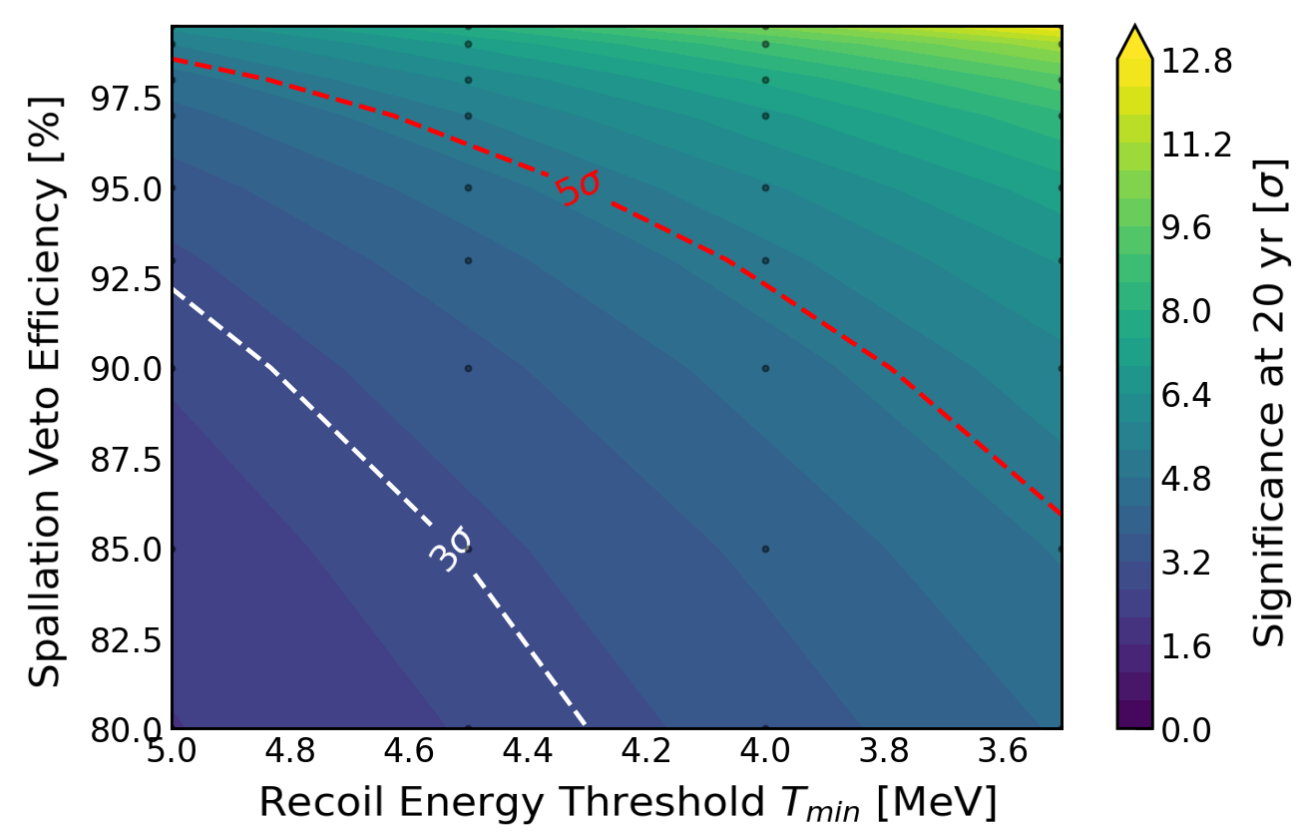}
    \end{minipage}\hfill
    \begin{minipage}[t]{0.48\linewidth}
        \centering
        \includegraphics[width=\linewidth, height=5.4cm]{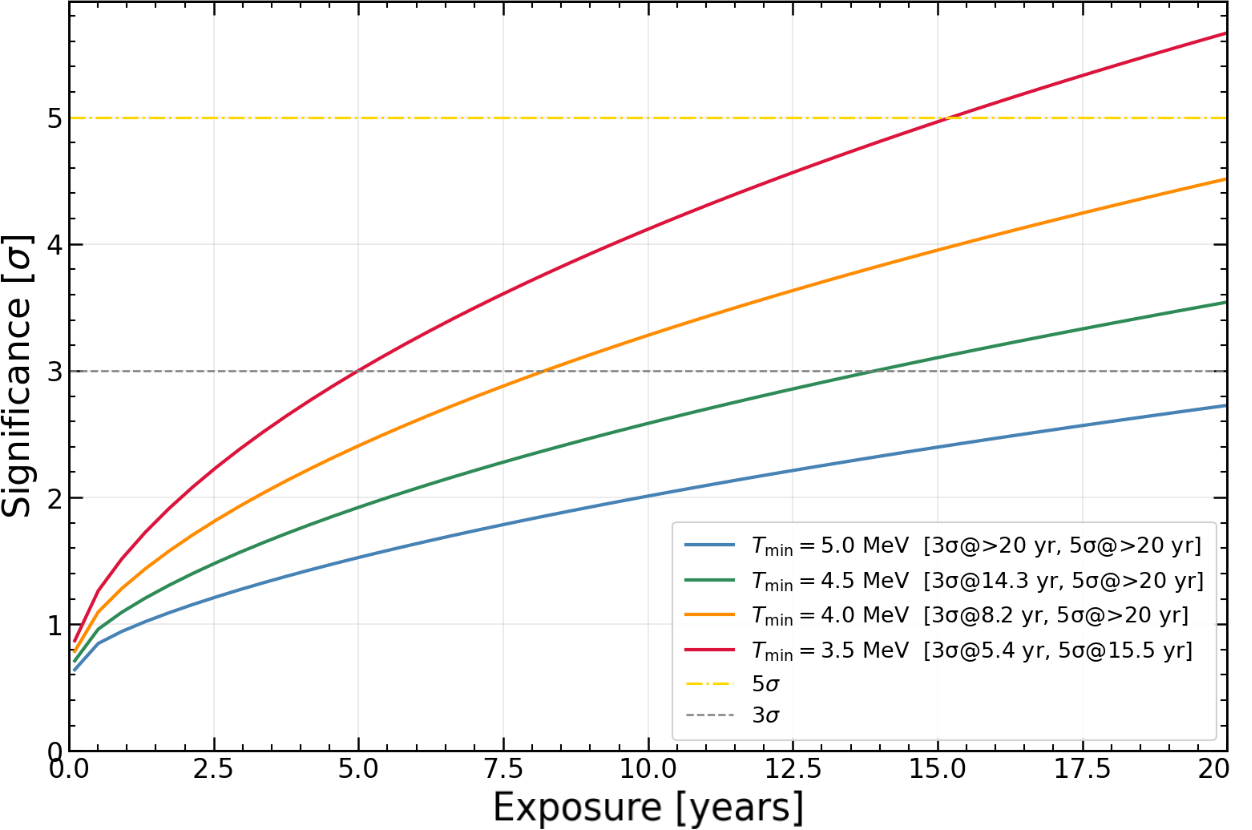}
    \end{minipage}
    \caption{\textit{Left}: Upturn discovery significance after 20 years of exposure in the plane of spallation veto efficiency and recoil energy threshold $T_{\rm min}$; the dashed white and red contours mark the $3\sigma$ and $5\sigma$ boundaries, respectively. \textit{Right}: upturn discovery significance vs.\ exposure for different recoil energy thresholds $T_{\rm min}=5.0,4.5,4.0,3.5$ MeV.}
    \label{fig:upturn_threshold_opt}
\end{figure}

Thus, given how strongly the residual spallation background limits the sensitivity, we further study the joint dependence of the achievable significance on the spallation veto efficiency and on the recoil energy threshold $T_{\rm min}$. Fig.~\ref{fig:upturn_threshold_opt} shows this optimization. The left panel presents the significance after 20 years in the two-dimensional plane of veto efficiency and $T_{\rm min}$; the importance of this information is based on the recent SK analysis for which a lower energy threshold has been adopted \cite{Super-Kamiokande:2023jbt}. Lowering the threshold from 5.0 to 3.5 MeV extends the accessible parameter space towards more moderate veto efficiencies while still reaching $3$--$5\sigma$, because more of the $^8$B spectrum enters the fit. However, lowering the threshold to 3.5~MeV introduces the radon-wall background (Sec.\ref{par:radon}), which is negligible above $\sim$5~MeV but grows rapidly below it. The net effect is still a modest gain in sensitivity, reaching 3--5$\sigma$ over the scanned range, but part of the benefit of a lower threshold is offset by this additional background. This threshold dependence is examined directly in the right panel, where the significance vs. exposure is shown for several values of $T_{\rm min}$ in a fixed veto scenario: lowering the threshold improves the sensitivity overall, with the $3\sigma$ level reached after 5.4, 8.2, and 14.3 years for $T_{\rm min}$=3.5, 4.0, and 4.5 MeV respectively, compared to beyond 20 years for $T_{\rm min}=5.0$ MeV; only the $T_{\rm min}=3.5$ MeV case reaches $5\sigma$ within the scanned exposure, at $\sim15.5$ years. We stress that lowering $T_{\rm min}$ improves the sensitivity mainly by capturing more of the $^8$B recoil-electron spectrum at the cross-section level, not by moving the fit window itself down to $T_{\rm min}$: the reconstructed-energy window is kept offset by 1.5~MeV above threshold in order to avoid the radon-wall background. The residual radon contamination remains negligible across the scanned range, confirming that the gain from a lower threshold is a genuine cross-section effect.

\section{The hep Neutrino Flux}
\label{sec:hep}

The $hep$ flux extends up to $\sim18.8$ MeV, above the $^8$B endpoint ($\sim15$ MeV), providing a spectrally distinct signature at the highest reconstructed energies accessible at ESSnuSB. We adopt the analysis window $E_{\rm rec}\in[14,25]$ MeV in 1 MeV bins, and use the same shape-driven, floating-normalization statistical treatment introduced in Sec.~\ref{sec:upturn}. The SSM predictions used are $\Phi_{^8\rm B}=4.21\times10^6$ cm$^{-2}$s$^{-1}$ and $\Phi_{\rm hep}=7.88\times10^3$ cm $^{-2}$ s $^{-1}$~\cite{Bahcall:2004pz} for energy-integrated flux values, giving an expected background of $^8$B-only of $\sim$1077 events/year in the analysis window, versus a combined $^8$ B $+hep$ (SSM) yield of $\sim$1204 events/year, i.e., \ a contribution of $hep$ of approximately 126 events/year.

\begin{figure}[h]
    \centering
    \includegraphics[width=0.65\linewidth, height=6.5cm]{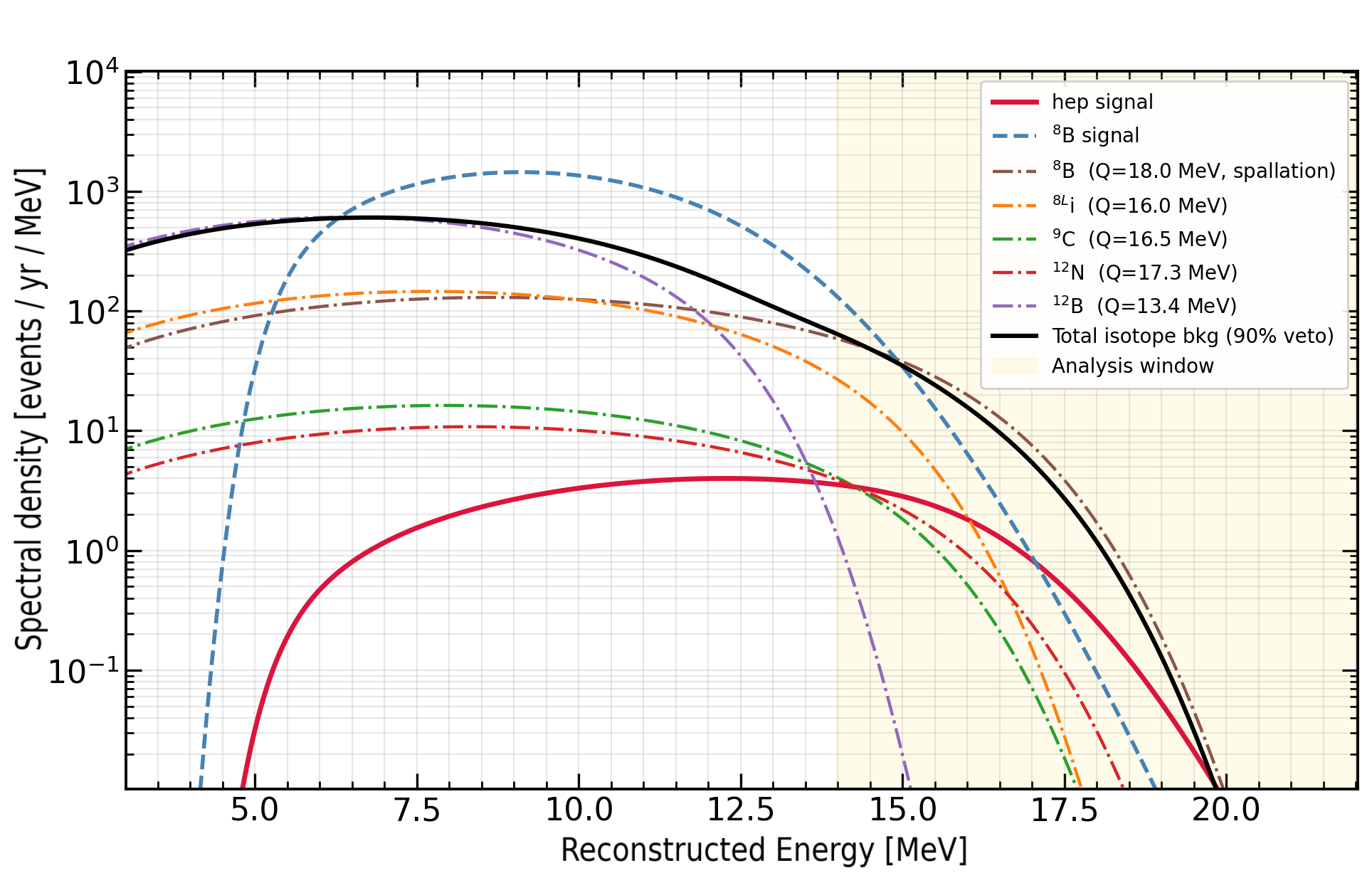}
    \caption{$hep$ signal (red) and $^8$B signal (blue dashed) compared to the individual isotope spallation backgrounds ($^8$B, $^8$Li, $^9$C, $^{12}$N, $^{12}$B, with their respective $\beta$-decay endpoints $Q$) and their sum after a 90\% veto (black). The shaded region marks the $hep$ analysis window. The full isotope-resolved decomposition is shown to justify why none of these backgrounds is dropped from the model: although a narrower window could exclude the lowest and highest endpoint isotopes, it would also discard a substantial fraction of the $hep$ flux itself, so the window is kept wide and the full background sum is retained instead.}
    \label{fig:hep_spectra}
\end{figure}

Fig.~\ref{fig:hep_spectra} shows the reconstructed spectra relevant to the \textit{hep} analysis, including the isotope-resolved spallation background introduced in Sec.~\ref{sec:solar}. The spallation isotopes ($^8$B, $^8$Li, $^9$C, $^{12}$N, $^{12}$B - the spallation $^8$B being a cosmogenic background distinct from the solar $^8$B signal) are shown together with their $\beta$-decay endpoints $Q$, alongside their total after a 90\% veto, the solar $^8$B background, and the \textit{hep} signal. While each individual isotope contribution falls rapidly above its own endpoint, their sum still exceeds the $hep$ signal by a sizeable margin throughout most of the analysis window. Above $\sim17$--18 MeV, close to the $hep$ end-point, $hep$ spectrum become comparable to or larger than the residual background;  this is the spectral region that drives the discriminating power of the analysis. The full isotope-resolved decomposition in Fig.~\ref{fig:hep_spectra} is shown to justify why no isotope is dropped from the background model: a narrower window could in principle exclude the lowest endpoint ($^{12}$B) and the highest-energy tails of the other isotopes, but only at the cost of discarding a significant fraction of the $hep$ signal itself. Therefore, the $[14,25]$~MeV window is kept wide. Furthermore, dropping its lower edge would recover some additional $hep$ events, but the background that would re-enter is not a small radon tail as in the $^8$B case - it is the $^8$B flux itself, $\sim500\times$ larger than $hep$, whose recoil spectrum is not negligible in that region.

\begin{figure}[h]
    \centering
    \begin{minipage}[t]{0.48\linewidth}
        \centering
        \includegraphics[width=\linewidth]{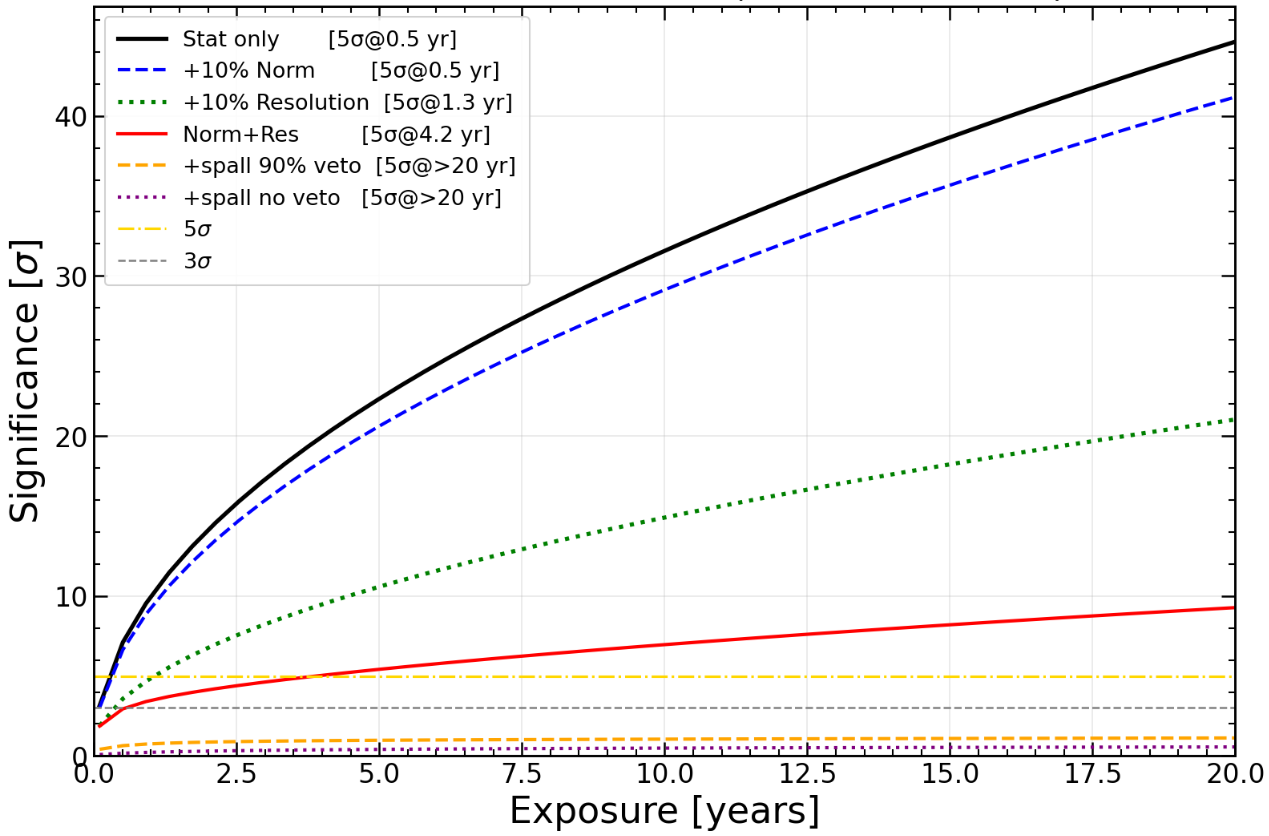}
    \end{minipage}\hfill
    \begin{minipage}[t]{0.48\linewidth}
        \centering
        \includegraphics[width=\linewidth]{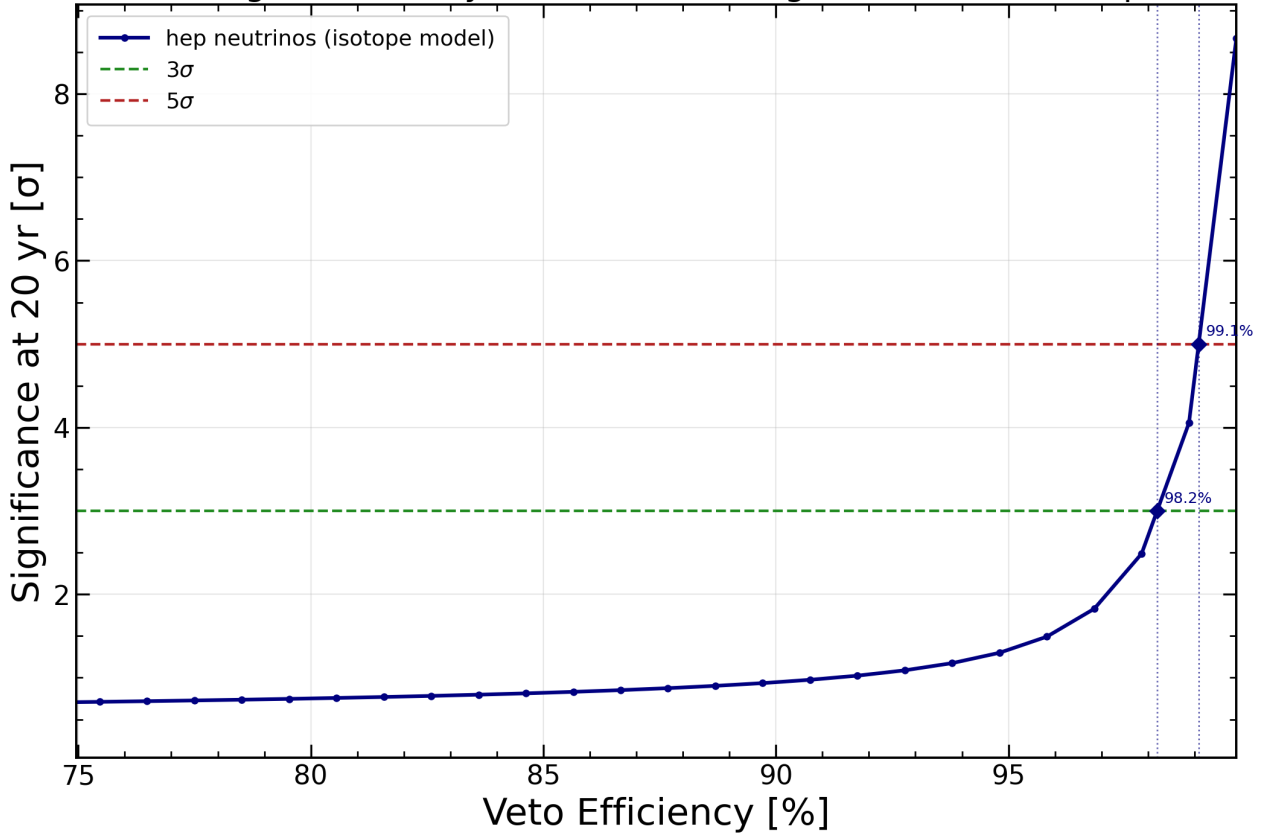}
    \end{minipage}
    \caption{\textit{Left}: $hep$ discovery significance vs.\ exposure: statistical only (black), $+10\%$ normalization (blue dashed), $+10\%$ resolution (green dotted), Norm+Res combined (red), and Norm+Res with a 90\% or no spallation veto (orange dashed, purple dotted). \textit{Right}: $hep$ discovery significance after 20 years of exposure as a function of the spallation veto efficiency, with the $3\sigma$ and $5\sigma$ reference lines shown.}
    \label{fig:hep_veto}
\end{figure}

To assess the discovery sensitivity, the null hypothesis $H_0$ includes only the $^8$B contribution, with its normalization free as in Sec.~\ref{sec:upturn}; the $H_1$ data includes both $^8$B and the SSM $hep$ flux. Fig.~\ref{fig:hep_veto} shows the resulting significance. In the purely statistical case, $5\sigma$ is reached within the first year of operation; adding normalization and resolution uncertainties delays this to $\sim4$ years, since a worse-than-expected energy resolution can redistribute $^8$B events into the $hep$-dominated bins and mimic part of the signal, while the spallation background is again the dominant limitation, keeping the significance well below $5\sigma$ within 20 years even with a 90\% veto. 

We scan the required spallation veto efficiency directly.  The left panel of the Fig.~\ref{fig:hep_veto} shows the significance after 20 years as a continuous function of veto efficiency, confirming that a rejection efficiency in excess of $\sim99\%$ is required before the significance rises sharply towards discovery level, for the case of 5 MeV threshold (notice that in this case the role of the threshold is not significant since the $hep$ neutrino spectrum lies well beyond the considered energy threshold). Below this value, the sensitivity increases only gradually with veto efficiency, mirroring the behaviour already observed for the upturn (Fig.~\ref{fig:upturn_sens}, right panel).

\begin{figure}[h]
    \centering
        \centering
        \includegraphics[width=0.5\linewidth, height =5.5cm]{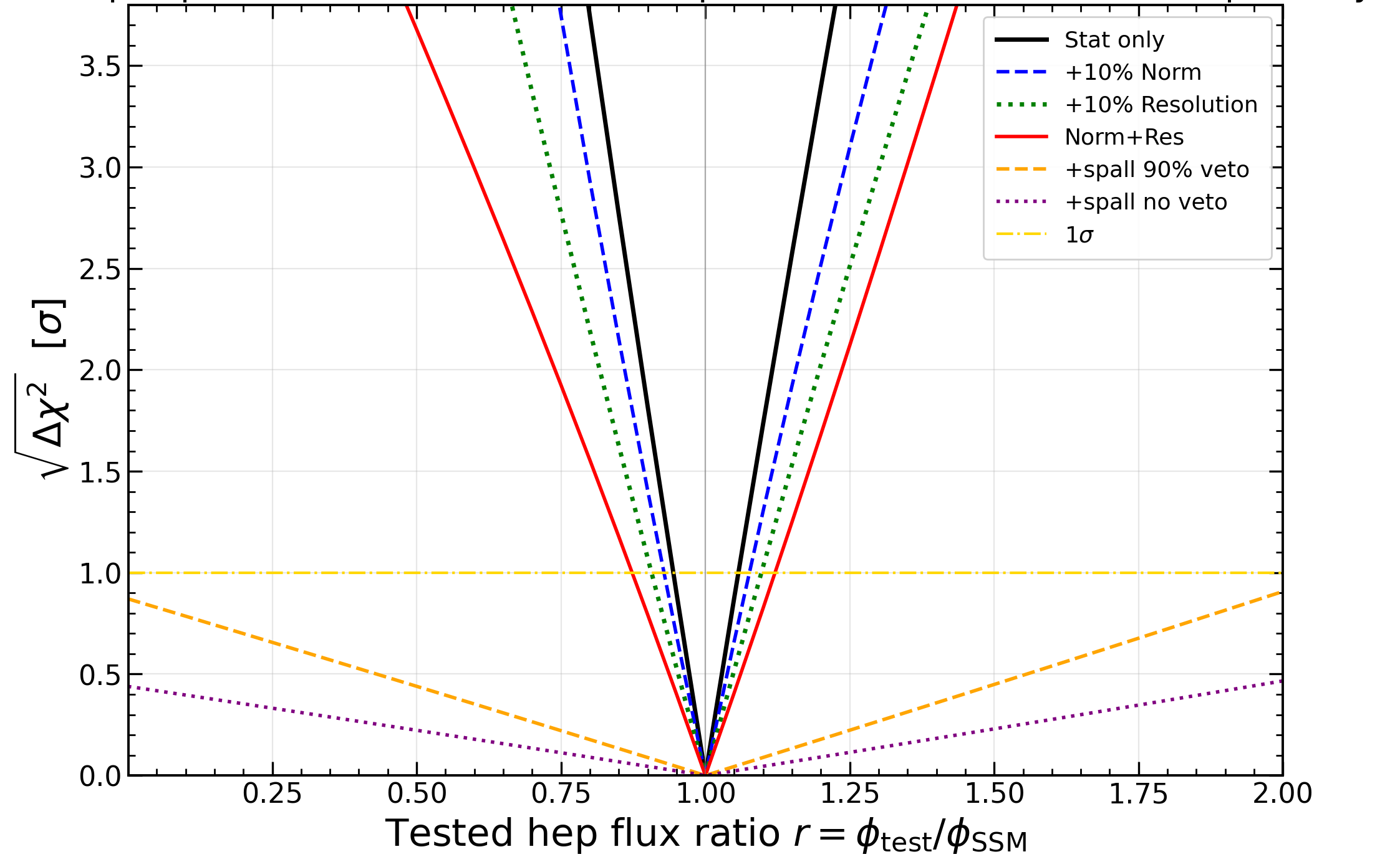}
    \caption{Precision measurement of the $hep$ normalization $r = \phi_{\rm hep}/\phi_{\rm hep}^{\rm SSM}$ for various scenarios with no background: statistical only (black), $+10\%$ normalization (blue dashed), $+10\%$ resolution (green dotted), Norm+Res combined (red), and Norm+Res with a 90\% or no spallation veto (orange dashed, purple dotted). The cases in which the spallation background is included are shown as orange and purple line for 90\% veto and no veto, respectively.}
    \label{fig:hep_prec}
\end{figure}

Beyond a binary discovery, we quantify in Fig. \ref{fig:hep_prec} the precision with which the hep flux normalization $r = \phi_{\rm hep}/\phi_{\rm hep}^{\rm SSM}$ can be measured for a fixed exposure, by scanning the profile likelihood as a function of $r_{\rm test}$ and computing $\sqrt{\Delta\chi^2} \equiv \sqrt{q(r) - q_{\rm min}}$.  In the purely statistical case a $\sim$6\% measurement of $r$ at $1\sigma$ is achievable after 10 years of exposure; adding normalization,
resolution, or both systematics widens this to 8\%, 10\%, and 12\%
respectively. Once the spallation background is included, the normalization is no longer meaningfully constrained at this exposure: with a 90\% veto, $r$ cannot be resolved to better than a factor of two at the $1\sigma$ level, highlighting again the importance of an extremely efficient background rejection to successfully detect and measure the \textit{hep} neutrino flux.

\section{Conclusions}

\label{sect:concl}
We have presented a sensitivity study of the ESSnuSB detector for two key solar neutrino physics goals: the MSW spectral upturn in the $^8$B survival probability, and the first direct detection of the $hep$ neutrino flux. Both analyses are designed as shape-driven measurements, with the overall event-rate normalization left free in the fit, so that the discovery power relies on the spectral distortion between hypotheses rather than on the absolute flux normalization. Without environmental backgrounds and under realistic detector systematic assumptions (10\% energy resolution and normalization uncertainties), both signals are, in principle, within reach on a timescale of a few years. However, once the cosmic-ray muon spallation background is included, the picture changes substantially: even a 90\% spallation veto efficiency, consistent with the performance historically achieved by Super-Kamiokande and with recent JUNO results, leaves both measurements well short of discovery level over a 20-year exposure.

Our dedicated scans of the spallation veto efficiency and of the recoil energy threshold show that this conclusion is not simply a matter of exposure: for both the upturn and the hep flux, the significance remains modest until the veto efficiency approaches its highest scanned values, with discovery-level ($5\sigma$) sensitivity for both analyses requiring $\sim$99\% rejection; for the upturn, evidence-level ($3\sigma$) sensitivity is reached somewhat earlier, around 94--95\%. For the upturn analysis, lowering the recoil energy threshold provides a complementary handle, recovering additional signal statistics at the cost of a larger low-energy background, and can partially compensate for a more moderate veto performance. These findings highlight three priorities for the ESSnuSB solar neutrino program: (i) achieving spallation rejection efficiencies significantly beyond the 90\% benchmark, which our results identify as the single most important requirement for both analyses; (ii) precise calibration of the detector energy resolution across the 5--25 MeV range, the most impactful systematic uncertainty besides the background; and (iii) a dedicated optimization of the recoil energy threshold, jointly with the veto efficiency, to maximize the physics reach. 

Detailed Monte Carlo simulations of the detector response, a Zinkgruvan site validation of the assumed per-isotope spallation rates and veto efficiencies (currently scaled from Super-Kamiokande), and dedicated studies of the day--night asymmetry (Appendix~\ref{app:dn}) remain important next steps for the ESSnuSB solar neutrino program.

\appendix
\section{Day--Night Asymmetry}
\label{app:dn}

\begin{figure}[htbp]
    \centering
    \includegraphics[width=\textwidth]{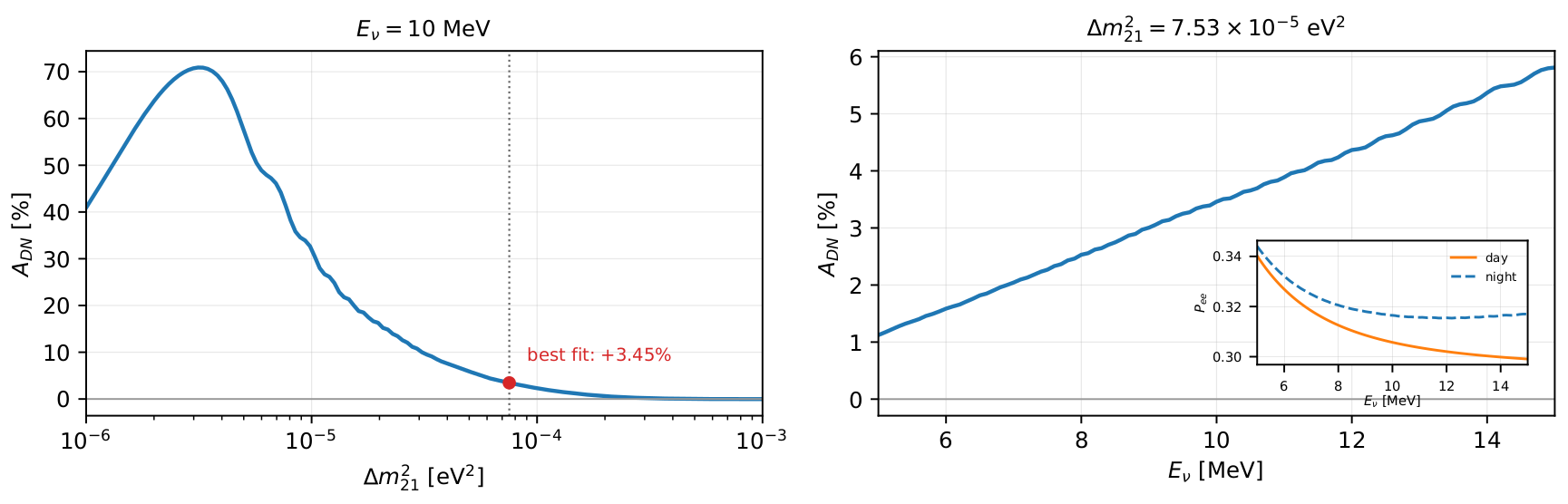}
    \caption{Annual-average day--night asymmetry of \(^{8}\)B solar neutrinos at the latitude of Zinkgruvan (\(59^\circ\)N). The left panel shows \(A_{DN}\) as a function of \(\Delta m^2_{21}\) for \(E_\nu=10~\mathrm{MeV}\). The right panel shows the energy dependence of the asymmetry for the same value of \(\Delta m^2_{21}\), while the inset displays the day-time and night-time electron-neutrino survival probabilities.}
    \label{fig:dn_asymmetry}
\end{figure}

A complementary probe of solar-neutrino oscillations is the day--night asymmetry,
\begin{equation}
A_{DN} = 2\frac{R_N-R_D}{R_N+R_D},
\end{equation}
where \(R_N\) and \(R_D\) denote the night-time and day-time event rates, respectively. This asymmetry arises from the partial regeneration of the electron-neutrino flavour when solar neutrinos propagate through the Earth during the night. Its magnitude depends on the neutrino energy and on the neutrino trajectories through the Earth, and therefore also on the geographic latitude of the detector. In particular, lower-latitude sites generally provide greater access to long trajectories crossing the denser regions of the Earth, whereas at high latitudes the distribution of night-time trajectories varies strongly throughout the year.
Fig.~\ref{fig:dn_asymmetry} shows the calculated day--night asymmetry for \(^{8}\)B solar neutrinos at the latitude of Zinkgruvan, approximately \(59^\circ\)N. The left panel shows its dependence on \(\Delta m^2_{21}\) at \(E_\nu=10~\mathrm{MeV}\), while the right panel shows its energy dependence for the adopted best-fit value \(\Delta m^2_{21}=7.53\times10^{-5}~\mathrm{eV}^2\). At \(E_\nu=10~\mathrm{MeV}\), the corresponding annual-average asymmetry is approximately \(3.4\%\).

For comparison, the corresponding day--night asymmetry expected at DUNE is approximately \(4.5\%\) for a similar value of \(\Delta m^2_{21}\)~\cite{Capozzi:2018dat}, and is larger than the value obtained for ESSnuSB. The DUNE value of Ref.~\cite{Capozzi:2018dat} is an event-rate asymmetry for $\nu_e$ charged-current interactions on argon, folded with the $^8$B spectrum, cross section, threshold and detector response, so it is not directly comparable with the fixed-energy probability asymmetry shown here. For the same quantity we find only a weak dependence on latitude between Zinkgruvan and the DUNE site (a few per cent in relative terms), and the larger DUNE value mainly reflects its higher effective energy threshold, since the asymmetry grows with energy. The results shown in Fig.~\ref{fig:dn_asymmetry} are averaged over a full year. However, at the high latitude of Zinkgruvan, both the duration of the night and the distribution of solar zenith angles vary considerably with the season. Consequently, the Earth-matter contribution may exhibit a seasonal dependence that is washed out in the annual average. Although the annual-average asymmetry is relatively modest, the large exposure expected at ESSnuSB may allow this seasonal variation to be investigated through a time-dependent analysis over a multi-year data-taking period.

\section*{Acknowledgments}

Funded by the European Union, Project 101094628. Views and opinions expressed are however those of the author(s) only and do not necessarily reflect those of the European Union. Neither the European Union nor the granting authority can be held responsible for them. We acknowledge further support provided by the following research funding agencies: Centre National de la Recherche Scientifique, France; Deutsche Forschungsgemeinschaft
Projektnummer 423761110, and under the Excellence Strategy of the Federal Government and the Länder, Germany; the European Union’s Horizon 2020 research and innovation programme under the Marie Skłodowska-Curie grant agreement No 860881-HIDDeN; the European Union NextGenerationEU, through the National Recovery and Resilience Plan of the Republic of Bulgaria, project No. BG-RRP-2.004-0008-C01; Roland Gustafssons Stiftelse för teoretisk fysik, Sweden; Croatian Science Foundation under the project numbers HRZZ-DOK-NPOO-2023-10-1262 and HRZZ-DOK-2025-02-2372; within the framework of the second Swiss Contribution under grant MAPS IZ11Z0$\_$230193 by the Swiss National Science Foundation (SNSF) and the Croatian Science Foundation (HRZZ); Ministry of Science, Education and Youth of the Republic of Croatia via grant No. PK.1.1.10.0002; as well as support provided by the universities and laboratories to which the authors of this report are affiliated, see the author list on the first page.

\bibliographystyle{JHEP}
\bibliography{bibliography}

\end{document}